\documentclass[showpacs,twocolumn]{revtex4}
\usepackage{graphicx}
\usepackage{dcolumn}
\usepackage{bm}
\usepackage{color}
\usepackage[squaren]{SIunits}
\usepackage{amssymb}
\usepackage{natbib}

\begin{document}

\newcommand{\ie}{{\it i.e.}}
\newcommand{\eg}{{\it e.g.}}
\newcommand{\etal}{{\it et al.}}


\title{Superconductivity at 32 K in single crystal Rb$_{0.78}$Fe$_2$Se$_{1.78}$}

\author{A. F. Wang, J. J. Ying, Y. J. Yan, R. H. Liu, X. G. Luo$^\dag$, Z. Y. Li,
X. F. Wang, M. Zhang, G. J. Ye, P. Cheng, Z. J. Xiang and  X. H. Chen}
\altaffiliation{E-mail of X.H.C: chenxh@ustc.edu.cn\\$^\dag$ E-mail of X.G.L: xgluo@mail.ustc.edu.cn}
\affiliation{Hefei National Laboratory for Physical Science at
Microscale and Department of Physics, University of Science and
Technology of China, Hefei, Anhui 230026, People's Republic of
China}

\date{\today}


\begin{abstract}

We successfully grew the high-quality single crystal of
Rb$_{0.78}$Fe$_2$Se$_{1.78}$, which shows sharp superconducting
transition in magnetic susceptibility and electrical resistivity.
Resistivity measurements show the onset superconducting transition
($T_{\rm c}$) at 32.1 K and zero resistivity at 30 K. From the
low-temperature iso-magnetic-field magnetoresistance, large upper
critical field $H_{\rm c2}$(0) has been estimated as high as 180 T
for in-plane field and 59 T for out-of-plane field. The anisotropy
$H^{ab}_{\rm c2}$(0)/$H^{c}_{\rm c2}$(0) is around 3.0, right lying
between those observed in K$_x$Fe$_2$Se$_2$ and Cs$_x$Fe$_2$Se$_2$.

\end{abstract}


\maketitle

The newly discovered iron-based superconductors have attracted
worldwide attention in past three
years\cite{Kamihara,chenxh,ZARen,rotter1,Liu} because of their high
superconducting transition temperature ($T_{\rm c}$ as high as 55 K)
and the fact that superconductivity emerges proximity to
magnetically ordered state.\cite{PCDai,hchen} The fact that
superconductivity in iron-pnictide compounds is closely related to
magnetic correlations inspires researchers tending to connect them
with the high-$T_{\rm c}$ cuprates, in which superconductivity is
realized by suppressing the antiferromagnetic Mott-insulating state,
and attempting to understand the superconducting mechanism in the
same theoretical scenario for the both families. Up to now, a
variety of Fe-based superconductors, such as ZrCuSiAs-type $Ln$FeAsO
($Ln$-1111, $Ln$ is rare earth
elements)~\cite{Kamihara,chenxh,ZARen}, ThCr$_2$Si$_2$-type
$Ae$Fe$_2$As$_2$ ($Ae$-122, $Ae$ is alkali earth
elements)~\cite{rotter1}, Fe$_2$As-type $A$FeAs ($A$-111, $A$ is Li
or Na)~\cite{CQJin, CWChu, Clarke} and anti-PbO-type Fe(Se,Te)
(11)\cite{MKWu}, have been discovered. Antiferromagnetic spin
density wave instability usually exists in the parent compound of
superconducting $Ln$-1111 and $Ae$-122, and even coexists with the
superconductivity in slightly doping levels of $Ln$-1111, $Ae$-122
and $A$-111. While for 11 phase, the magnetism is quite complicated
and its relationship to superconductivity remains more unrecognized.

All of the above mentioned Fe-base superconductors have a common
structural feature with the edge-sharing FeAs$_4$ (FeSe$_4$)
tetrahedra forming FeAs (FeSe) layers. The superconductivity in
these compounds is thought to be intimately associated with the
height of anion from Fe layer~\cite{Mizuguchi1}. FeAs-based
compounds usually possess cations or building block between the FeAs
layers, while Fe(Se,Te) family has an extremely simple structure with
only FeSe layers stacking along {\sl c}-axis without other cations
between them.\cite{MKWu}  High pressure has been used to change the
height of anion from Fe layer in Fe(Se,Te). Especially, $T_{c}$ can
reach 37 K (onset) under 4.5 GPa from  8 K in FeSe\cite{cava} with a
pressure dependent ratio of $T_{\rm c}$ as large as d$T_{\rm
c}$/d$P$ $\sim$ 9.1 K/GPa, which is the highest pressure effect
among all the Fe-base superconductors.\cite{cava} Tl has been
attempted to intercalate into between the FeSe layers to change the
local structure of FeSe family. However, an antiferromagnetic
ordering forms at the temperature as high as 450 K,\cite{YingJJ} and
no superconductivity is observed in TlFe$_2$Se$_2$. Very recently, the
alkali atoms K and Cs are successfully intercalated into between the
FeSe layers, and superconductivity has been enhanced from $T_{\rm
c}$ = 8 K of pure FeSe to 30 K and 27 K (onset) without any external
pressure.\cite{xlchen,Mizuguchi,Krzton,YingK} It indicates that
$T_{\rm c}$ in FeSe family can really be enhanced by intercalating
cations into between the FeSe layers.

In this communication, we successfully grew the single crystals of a
new superconductor Rb$_x$Fe$_2$Se$_2$ by using self-flux method. The
crystals showed the onset $T_{\rm c}$ of 32.1 K and zero resistivity
at about 30 K. Nearly 100$\%$ superconducting volume fraction was
observed through the zero-field-cooling (ZFC) magnetic
susceptibility measurements. Upper critical field $H_{\rm c2}$(0)
was estimated from iso-magnetic-field magnetoresistance as high as
180 T with field applied in {\sl ab}-plane and 59 T with field
applied along {\sl c}-aixs.

\begin{figure}
\includegraphics[width = 0.45\textwidth]{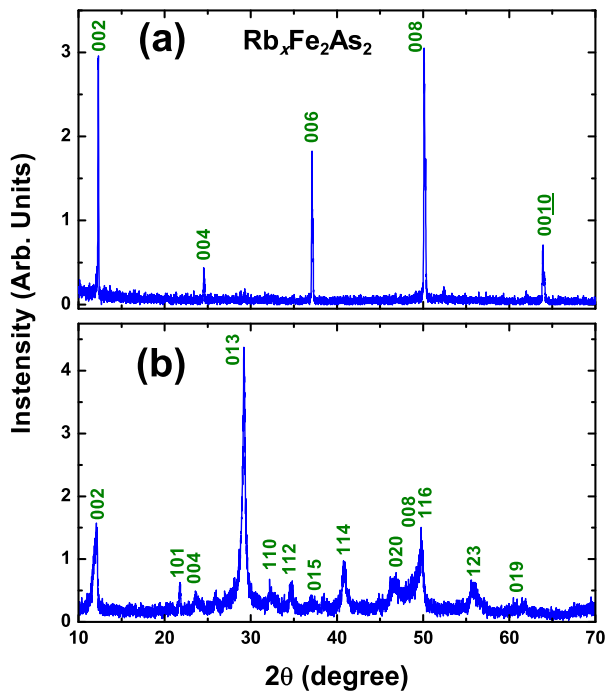}
\caption{(Color online) X-ray diffraction patterns for
Rb$_x$Fe$_2$Se$_2$, (a): The single crystal X-ray diffraction
pattern; (b): X-ray diffraction pattern of the powdered
Rb$_x$Fe$_2$Se$_2$.}
\end{figure}

Single crystals Rb$_x$Fe$_2$Se$_2$ were grown by self-flux method.
Starting material FeSe was obtained by reacting Fe powder with Se
powder with Fe: Se = 1: 1 at 700$\celsius$ for 4 hours. Rb pieces
and FeSe powder were put into a small quartz tube with nominal
composition of Rb$_{0.8}$Fe$_2$Se$_2$. Due to the high activity of
Rb metal, the single wall quartz tube will be corrupted and broken
during the growth procedure. Therefore, double wall quartz tube is
used here. The small quartz tube was sealed under high vacuum, and
then was put in a bigger quartz tube following by evacuating and
being sealed. The mixture was heated to 980 $\celsius$ in 10 hours
and kept for 4 hours, and then melt at 1080 $\celsius$ for 2 hours,
and later slowly cooled down to 780 $\celsius$ with 6
$\celsius$/hour. After that, the temperature  was cooled down to
room temperature by shutting down the furnace. The obtained single
crystals show the flat shiny surface with dark black color. The
crystals are easy to cleave and thin crystals with thickness less
than 100 $\mu$m can be easily obtained.

The single crystals were characterized by X-ray diffraction (XRD),
Energy dispersive X-ray (EDX) spectroscopy, magnetic susceptibility,
and electrical transport measurements. Powder XRD and single crystal
XRD were performed on TTRAX3 theta/theta rotating anode X-ray
Diffractometer (Japan) with Cu K$\alpha$ radiation and a fixed
graphite monochromator. Magnetic susceptibility measurements were
carried out using the {\sl Quantum Design} MPMS-SQUID. The
measurement of resistivity and magnetoresistance were done on the
{\sl Quantum Design} PPMS-9.

Figure 1 shows the X-ray single crystal diffraction (Fig. 1a) and
powder XRD (Fig. 1b) after grounding the single crystals into
powder. Only (00$l$) reflections were recognized in Fig. 1a,
indicating that the crystals of Rb$_x$Fe$_2$Se$_2$ were perfectly
grown along {\sl c}-axis. From the powder XRD patterns in Fig. 1b,
the lattice constants were calculated based on the symmetry I4/mmm
with lattice parameters $a$ = 3.925 ${\rm \AA}$ and $c$ = 14.5655
${\rm \AA}$. Lattice constants of $a$ and $c$ lie between those of
K$_x$Fe$_2$Se$_2$ and Cs$_x$Fe$_2$Se$_2$, respectively. It is
consistent with the expectation based on variation of the radius of
the K, Rb, Cs ions (K 1.51${\rm \AA}$, Rb 1.63${\rm \AA}$, Cs
1.78${\rm \AA}$).\cite{Shannon} The actual compositions of the crystals were
determined by EDX using an average of different 4 points. It is
found that the composition is homogeneous in the crystals. The
actual composition is Rb: Fe: Se = 0.78: 2: 1.78, indicating the
existence of deficiencies at K sites and Se sites. Such deficiency
is similar to our previous report for K$_x$Fe$_2$Se$_2$
~\cite{YingK}, but is sharply in contrast to other reports of both K
and Fe deficiencies in K$_x$Fe$_2$Se$_2$ and
Cs$_x$Fe$_2$Se$_2$.\cite{xlchen,Mizuguchi,Krzton}

\begin{figure}[ht]
\includegraphics[width = 0.45\textwidth]{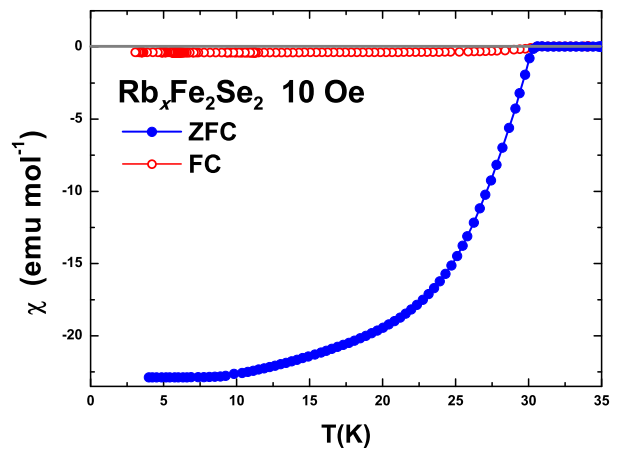}
\caption{(Color online) Temperature dependence of the Zero-field
cooling and field cooling susceptibility taken at 10 Oe with the
magnetic field parallel to the {\sl ab}-plane for the single crystal
Rb$_{0.78}$Fe$_2$Se$_{1.78}$.}
\end{figure}

Figure 2 shows magnetic susceptibility as a function of temperature
below 35 K for single crystal Rb$_{0.78}$Fe$_2$Se$_{1.78}$ under a
magnetic field of 10 Oe. The zero-field-cooling (ZFC) and field
cooling (FC) susceptibilities show that the superconducting shield
begins to emerge at about 30.6 K and then show a sharp transition.
The ZFC magnetic susceptibility becomes saturation below 10 K,
indicating high quality of single crystal. The superconducting
volume fraction estimated from the ZFC magnetization at 4 K is
100$\%$. All of these demonstrate a bulk superconductivity nature in
Rb$_{0.78}$Fe$_2$Se$_{1.78}$ single crystals.

\begin{figure}[ht]
\includegraphics[width = 0.45\textwidth]{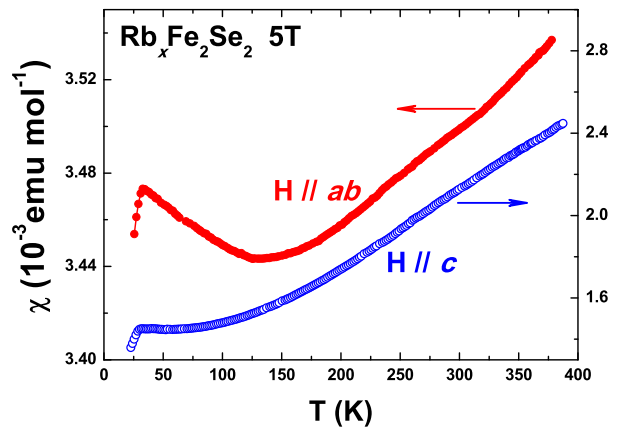}
\caption{(Color online) The magnetic susceptibility at 5 T for
single crystal Rb$_{0.78}$Fe$_2$Se$_{1.78}$ with the magnetic field
along and perpendicular to {\sl c}-axis.}
\end{figure}

Figure 3 shows the magnetic susceptibility of
Rb$_{0.78}$Fe$_2$Se$_{1.78}$ with the magnetic field of 5 T applied
parallel and perpendicular to the {\sl c}-axis from 10 K to 400 K.
At low temperature, superconducting trace can still be found because
of a drop of susceptibility. When magnetic field was applied along
c-axis, the magnetic susceptibility gradually decreases with
decreasing  the temperature. The susceptibility shows a minimum at
about 120 K with the magnetic field applied within ab-plane. Above
120 K, the susceptibility monotonically increases with increasing
temperature; while gradually increases with decreasing temperature
down to about 40 K just above superconducting transition
temperature. Although the in-plane $\chi(T)$ shows a minimum 120 K
above $T_{\rm c}$, the magnitude of the susceptibility only changes
by less than 2.5 $\%$ in the temperature range from 40 K to 400 K.
Such behavior of susceptibility in Rb$_{0.78}$Fe$_2$Se$_{1.78}$ is
exactly the same as that observed in
Cs$_{0.86}$Fe$_{1.66}$Se$_{2}$.\cite{YingK} Therefore, such peculiar
behavior of susceptibility is common feature. The continuous
decrease of susceptibility with decreasing the temperature suggests
a strong antiferromagnetic spin fluctuation. Such spin fluctuation
could be related to the superconductivity.

\begin{figure}[ht]
\includegraphics[width = 0.45\textwidth]{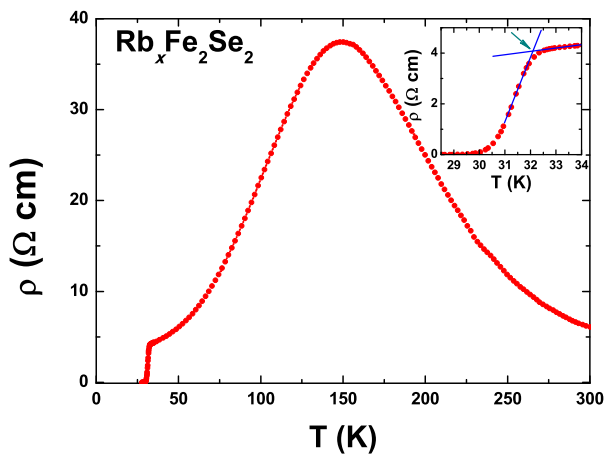}
\caption{(Color online)Temperature dependence of resistivity for
single crystal Rb$_{0.78}$Fe$_2$Se$_{1.78}$. The inset is the zoom
plot of resistivity around superconducting transition.}
\end{figure}

Figure 4 shows the in-plane resistivity as the function of
temperature for the Rb$_{0.78}$Fe$_2$Se$_{1.78}$. The
Rb$_{0.78}$Fe$_2$Se$_{1.78}$ shows the semiconductor-like behavior
at the high temperature, and displays a maximum resistivity at about
150 K, and shows a metallic behavior below 150 K and a
superconducting transition at about 32 K. Similar resistivity has
been observed in K$_x$Fe$_2$Se$_2$.\cite{xlchen,YingK} It seems that
the resistivity behavior observed here is common feature. The
temperature corresponding to the maximum resistivity in
Rb$_{0.78}$Fe$_2$Se$_{1.78}$ is higher than that for
K$_x$Fe$_2$Se$_2$ reported by Guo {\it et al.} (around 100
K)~\cite{xlchen} and by Ying {\it et al.}(around 120
K)~\cite{YingK}, while less than that reported by Mizuguchi {\etal}
($\sim$200 K).\cite{Mizuguchi} The maximum resistivity in
Rb$_{0.78}$Fe$_2$Se$_{1.78}$ crystal here ($\sim$37 $\Omega$ cm) is
much larger than that of K$_x$Fe$_2$Se$_2$ in previous report
($\sim$3 $\Omega$ cm).\cite{Mizuguchi}  The temperature of the
maximum resistivity strongly depends on the sample. The different
temperature of the maximum resistivity could arise from the
vacancies at Fe or Se sites. The residual resistance ratio between
150 K and 33 K is as large as 9. With further decreasing the
temperature, superconductivity emerges at about 32.1 K and
resistivity reaches zero at around 30 K. These values are very close
to those observed in K$_x$Fe$_2$Se$_2$.\cite{xlchen,Mizuguchi} The
resistivity of Rb$_{0.78}$Fe$_2$Se$_{1.78}$ crystal are 6 $\Omega$
cm at room temperature, which is much larger than those of FeSe
single crystals \cite{Braithwaite} and the other iron-pnictide
superconductors\cite{wang}. This may arise from the large disorder
induced by deficiencies of Fe or Se. Occurrence of superconductivity
in a system with so high resistivity demands further theoretical and
experimental investigation.

\begin{figure}[ht]
\includegraphics[width = 0.45\textwidth]{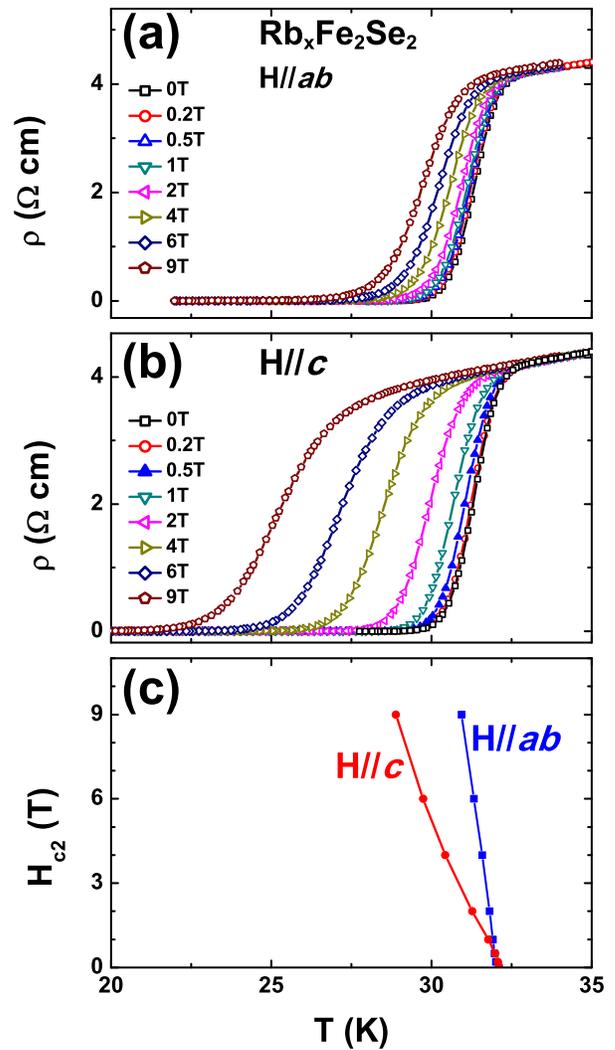}
\caption{(Color online) (a) and (b) show the temperature dependence
of resistivity for Rb$_{0.78}$Fe$_2$Se$_{1.78}$ with the magnetic
field parallel and perpendicular to the {\sl ab}-plane,
respectively; (c): The temperature dependence of upper critical
field $H_{\rm c2}$(T) for Rb$_{0.78}$Fe$_2$Se$_{1.78}$.}
\end{figure}

Resistivity as a function of temperature under the magnetic field
applied in {\sl ab}-plane and along the {\sl c}-axis is shown in
Fig. 5a and 5b. The transition temperature of superconductivity is
suppressed gradually and the transition is broadened with increasing
the magnetic field. Obvious difference for the effect of field along
different direction on the superconductivity can observed. In order
to study this difference clearly, we defined the $T_{\rm c}$ as the
temperature where the resistivity was 90$\%$ drop right above the
superconducting transition. The anisotropic $H_{\rm c2}$($T$) are
shown in Fig. 5c for the two field directions, respectively. Within
the weak-coupling BCS theory, the upper critical field at $T$=0 K
can be determined by the Werthamer-Helfand-Hohenberg (WHH)
equation\cite{Werthamer} $H_{\rm c2}(0)=0.693[-(dH_{\rm
c2}/dT)]_{T_{\rm c}}T_{\rm c}$. From Fig. 5c, we can have $[-(dH_{\rm
c2}^{ab}/dT)]_{T_{\rm c}}$ = 8.09 T/K, $[-(dH_{\rm
c2}^{c}/dT)]_{T_{\rm c}}$ = 2.66 T/K and $T_{\rm c}$ = 32.1 K. Then
the $H_{\rm c2}(0)$ can be estimated to be 180 T and 59 T with the
magnetic field applied in {\sl ab}-plane and along the {\sl c}-axis,
respectively. These values are less than  that in K$_x$Fe$_2$Se$_2$
\cite{Mizuguchi,YingK}, while larger than that in
CsFe$_2$Se$_2$.\cite{YingK}  The anisotropy $H^{ab}_{\rm
c2}$(0)/$H^{c}_{\rm c2}$(0) is about 3.0 and this value just lies
right between K$_x$Fe$_2$Se$_2$ and Cs$_x$Fe$_2$Se$_2$.   This
anisotropy value is larger than 1.70$\sim$1.86 in
Ba$_{0.60}$K$_{0.40}$Fe$_2$As$_2$\cite{wenhh}, while less than
4$\sim$6 in F-doped NdFeAsO\cite{Jia}.

We have systematically grown the single crystals $A_x$Fe$_2$Se$_2$ ($A$ = K,
Cs and Rb). It is found that there exist some common features in
resistivity and magnetic properties for these crystals. A maximum
resistivity as shown in Fig. 4 is widely observed in
K$_x$Fe$_2$As$_2$~\cite{xlchen,YingK,Mizuguchi} and Rb$_x$Fe$_2$As$_2$.
Another common feature is that peculiar behavior of normal state
susceptibility as shown in Fig.3 is widely observed in
Cs$_x$Fe$_2$As$_2$~\cite{YingK} and Rb$_x$Fe$_2$As$_2$. It is found based on
the observation in Fig. 3 and 4 that the maximum resistivity nearly
coincides with the minimum susceptibility with magnetic field
applied within {\sl ab}-plane. It suggests that there exists a correlation
between the maximum resistivity and the minimum in-plane
susceptibility. It should be addressed that the deficiency of Fe and
Se is related to the ionic radius of alkali metals K, Rb and Cs. The
actual compositions of superconducting crystals are
K$_{0.86}$Fe$_2$Se$_{1.82}$\cite{YingK}, Rb$_{0.78}$Fe$_2$Se$_{1.78}$
and Cs$_{0.86}$Fe$_{1.66}$Se$_{2}$. It indicates that the vacancy in
conducting FeSe layers changes from Se site to Fe site with
increasing the ionic radius of alkali metals from K to Cs. It is
found that normal state resistivity and susceptibility strongly
depend on the vacancy in conducting FeSe layers. Further study on
the origin of the deficiency of Fe and Se should be required to
understand the normal state behavior, even the superconductivity of
$A_x$Fe$_2$Se$_2$ materials.

In conclusion, we successfully grew a new superconductor
Rb$_{0.78}$Fe$_2$Se$_{1.78}$ single crystals. $T_{\rm c}^{\rm
onset}$ is 32.1 K determined by resistivity measurement and zero
resistivity is reached at 30 K. The ZFC dc magnetic susceptibility
indicates that the crystal is fully diamagnetic. The large $H_{\rm
c2}$(0) is observed, being similar to that in other the
iron-pnictide superconductors\cite{Yuan}. The anisotropy
$H^{ab}_{\rm c2}$(0)/$H^{c}_{\rm c2}$(0) is 3.0, right lying between
those of K$_x$Fe$_2$Se$_2$ and Cs$_x$Fe$_2$Se$_2$. A common peculiar
susceptibility at the normal state is observed in
Rb$_{0.78}$Fe$_2$Se$_{1.78}$.
\\

{\bf ACKNOWLEDGEMENT} This work is supported by the Natural Science Foundation
of China and by the Ministry of Science and Technology of China,
and by Chinese Academy of Sciences.\\

\end{document}